# Magnitude to luminance conversions and visual brightness of the night sky

Salvador Bará[1,*], Martin Aubé[2], John Barentine[3,4], and Jaime Zamorano[5]

[1] *Departamento de Física Aplicada, Universidade de Santiago de Compostela, 15782 Santiago de Compostela, Galicia*
[2] *Physics department, Cégep de Sherbrooke, Canada*
[3] *International Dark-Sky Association, Tucson, AZ, 85719 USA*
[4] *Consortium for Dark Sky Studies, University of Utah, Salt Lake City, Utah 84112 USA*
[5] *Dept. Física de la Tierra y Astrofísica, Instituto de Física de Partículas y del Cosmos (IPARCOS), Universidad Complutense de Madrid, 28040 Madrid, Spain*

* E-mail: salva.bara@usc.gal

**ABSTRACT**

The visual brightness of the night sky is not a single-valued function of its brightness in other photometric bands, because the transformations between photometric systems depend on the spectral power distribution of the skyglow. We analyze the transformation between the night sky brightness in the Johnson-Cousins *V* band ($m_V$, measured in magnitudes per square arcsecond, mpsas) and its visual luminance (*L*, in SI units cd m$^{-2}$) for observers with photopic and scotopic adaptation, in terms of the spectral power distribution of the incident light. We calculate the zero-point luminances for a set of skyglow spectra recorded at different places in the world, including strongly light-polluted locations and sites with nearly pristine natural dark skies. The photopic skyglow luminance corresponding to $m_V$=0.00 mpsas is found to vary between 1.11-1.34 x 10$^5$ cd m$^{-2}$ if $m_V$ is reported in the absolute (AB) magnitude scale, and between 1.18-1.43 x 10$^5$ cd m$^{-2}$ if a Vega scale for $m_V$ is used instead. The photopic luminance for $m_V$=22.0 mpsas is correspondingly comprised between 176 and 213 μcd m$^{-2}$ (AB), or 187 and 227 μcd m$^{-2}$ (Vega). These constants tend to decrease for increasing correlated color temperatures (CCT). The photopic zero-point luminances are generally higher than the ones expected for blackbody radiation of comparable CCT. The scotopic-to-photopic luminance ratio (S/P) for our spectral dataset varies from 0.8 to 2.5. Under scotopic adaptation the dependence of the zero-point luminances with the CCT, and their values relative to blackbody radiation, are reversed with respect to photopic ones.

# 1. INTRODUCTION

Determining the conversion factors between the visual brightness of the night sky measured in SI luminance units (cd m$^{-2}$), and its brightness in magnitudes per square arcsecond (mpsas) in any given astrophysical photometric band is an issue of great interest for a variety of skyglow studies. A wide set of photometric systems are available for use in modern astronomy (Bessell, 2005). The theoretical modelling and observational measurements of the visual effects of the skyglow, both at light polluted sites and at pristine dark locations, have been traditionally carried out (and still are) in the classical Johnson-Cousins *V*-band (Bessel, 1990). The reasons for this choice can be probably traced back to the coarse similarity between this band and the human eye photopic sensitivity function, such that the results obtained in *V* could be taken as a first approximation to the ones that would correspond to human vision. However the Johnson-Cousins *V*-band is neither coincident with the photopic nor the scotopic eye sensitivity bands (CIE, 1990, 2010), and hence any reasonable estimation of the actual visual brightness of the night sky as perceived by a human observer requires of specific conversions. A widely used transformation between Johnson-Cousins *V* mpsas and visual brightness in nanoLamberts (nL) has been proposed by Garstang (1986, 1989), based on previous results by Allen (1973), and is still routinely applied in many works. However, this transformation is strictly valid only for blackbody radiation of a specific temperature (Bará, 2019). As a matter of fact, no single zero-point luminance with universal validity can be assigned to such transformation, because it depends not only on the particular definition of the intervening photometric bands (Bessell, 2005; Carroll & Ostlie, 2007) but also, and very significantly, on the spectral composition of the incoming light.

In previous works we developed a consistent definition of an absolute mpsas system for the human visual system (Bará, 2017), and reported the conversion factors for determining the luminance of blackbody radiation of arbitrary temperature as a function of its mpsas brightness in the Johnson-Cousins *B*, *V*, and *R* photometric bands (Bará, 2019).

Artificial and natural skyglow, however, seldom present a blackbody spectrum. This is a particularly relevant issue as the world continues its transition toward outdoor lighting sources with non-thermal spectra (Sánchez de Miguel et al, 2017). In this work we extend our previous results by calculating the equivalence between brightness in magnitudes per square arcsecond in the Johnson-Cousins *V* band ($m_V$) and visual luminance (*L*) in cd m$^{-2}$ for typical night skyglow spectra. The main aim of this work is to provide the detailed expressions that may allow the interested reader to carry out the appropriate transformations for their particular spectra, and to give some general insights about the expected value of the conversion factors for actual light polluted skies in urban and rural settings. Some examples of pristine dark skies are also included for comparison.

In sections 2 and 3 of this paper we briefly revisit some basic concepts of radiometry, as well as of visual and astrophysical photometry. Those readers with a basic knowledge of these issues may wish to skip these sections and go directly to the results in Section 4 "Transforming magnitudes-per-square-arcsecond to luminances". The approach described in Sections 2 to 4 has general validity and can be applied to convert to the CIE visual band the magnitudes modelled or observed in any astrophysical photometric band, not only those belonging to the Johnson-Cousins system. Numerical results are provided in Section 5 for the Johnson-Cousins *V* band due to its widespread use in the visual skyglow research.

## 2. LUMINANCE AND VISUAL PERCEPTION

Luminance is the basic physical quantity associated with the human percept of brightness. It is expressed in cd m$^{-2}$, the cd (*candela*) being the only base unit of the International System (SI) (BIPM, 2019) that crucially depends on the particular features of the human visual system. The cd=lm sr$^{-1}$ is the unit for luminous intensity, a measure of the amount of light (unit lm, *lumen*) propagating in a given direction per unit solid angle (unit sr, *steradian*). A related quantity, also important for vision applications, is illuminance, measured in lx=lm m$^{-2}$ (unit lx, *lux*), that is, the amount of light per unit surface area. But both luminance and illuminance turn out to depend on an even more disaggregated physical quantity: the spectral radiance.

The spectral radiance $L(\mathbf{r}, \boldsymbol{\alpha}; \lambda)$, or $L(\lambda)$ in short, is an intrinsic property of the electromagnetic field. It can be defined for every point of the space ($\mathbf{r}$) and for every direction of propagation ($\boldsymbol{\alpha}$), as:

$$L(\lambda) = \frac{d\Phi(\lambda)}{\cos(z) dA d\omega d\lambda}, \qquad (1)$$

(see Figure 1), that is, as the amount of radiant power $\Phi(\lambda)$, also known as 'radiant flux' (units W), propagating along that direction, per unit solid angle $d\omega$ (sr), per unit projected surface $dA_0 = \cos(z) dA$ (m$^2$), and per unit spectral interval $d\lambda$ (nm). Let us recall that directions in space can be specified in a sperical coordinate system by means of the polar angle ($z$) and the azimuth ($\phi$), which can be grouped together to form the direction vector $\boldsymbol{\alpha} = (z, \phi)$. The spectral radiance is measured in W m$^{-2}$ sr$^{-1}$ nm$^{-1}$ if the spectra are expressed as a function of the wavelength as in Eq.(1) but can also be given in W m$^{-2}$ sr$^{-1}$ Hz$^{-1}$ if the spectra are expressed as a function of the electromagnetic field frequency (Hz) instead.

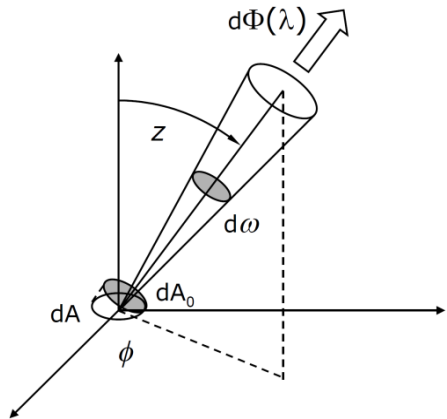

**Figure 1.** Geometry for the definition of spectral radiance

The spectral radiance is a physical quantity of practical importance for three main reasons: (a) it is directly related to the amount of power that is captured by any measurement instrument (including the human eye), (b) it allows the unambiguous calculation of any other radiometric or photometric quantity of interest, and (c) it determines the brightness (luminance) of every point of the retinal image as a consequence of the two reasons mentioned above.

As an example, let us consider the relationship between the spectral radiance and another widely used radiometric quantity, the spectral irradiance $E(\mathbf{r}; \lambda)$, or $E(\lambda)$ in short-hand notation. The spectral irradiance is the amount of power arriving to (or leaving) $\mathbf{r}$ per unit surface and per unit spectral interval. Note that this surface does not have to be that of a physical body: the irradiance can be defined at any point of the radiant field, as far as an elementary plane surface can be imagined containing it. The spectral irradiance is relevant for light pollution studies because astronomical magnitudes are just a way of reporting spectral irradiances integrated in the photometric observation passbands and expressed in a negative logarithmic scale. Thus, whereas the irradiance is the amount of radiant power per unit surface at $\mathbf{r}$, the radiance accounts for how that irradiance is built up from the contributions of power arriving to $\mathbf{r}$ from different directions. To calculate the spectral irradiance one just has to add up the spectral radiance arriving to the surface from all possible directions of its front-facing hemisphere $\Omega$, according to (McCluney, 2014):

$$E(\mathbf{r}; \lambda) = \int_\Omega L(\mathbf{r}, \boldsymbol{\alpha}; \lambda)\cos(z)\mathrm{d}\omega , \qquad (2)$$

where $\mathrm{d}\omega = \sin(z)\,\mathrm{d}z\mathrm{d}\phi$ is the elementary solid angle around the direction $\boldsymbol{\alpha} = (z, \phi)$, the polar angle $z$ is measured from the normal to the surface at $\mathbf{r}$, and the integration is extended to all directions in the $\Omega$ hemisphere, spanning a solid angle of $2\pi$ sr. The factor $\cos(z)$ in the integrand appears because in the proper definition of radiance (Eq. 1), the 'per unit surface' refers to a planar surface oriented perpendicularly to the direction of propagation, $\boldsymbol{\alpha}$.

On the other hand, as already pointed out, the spectral radiance allows one to calculate the amount of power entering the eye from every point of the environment along each direction of the visual field. The eye only captures that fraction of the flux emitted from $\mathbf{r}$ that is contained within a cone of directions whose vertex is $\mathbf{r}$ and its base is the size of the eye's pupil opening. So, not all radiant flux leaving $\mathbf{r}$ per unit surface reaches the eye retina. The object irradiance, per se, is not an adequate quantity to measure the retinal exposure. What determines the irradiance on the retina, and consequently the amount of power received by each photoreceptor cell, is the object radiance.

Retinal photoreceptor cells are not equally sensitive to all wavelengths. The three types of cone cells (S, L and M), the rod cells, and the intrinsically photosensitive retinal ganglion cells (ipRGCs), have specific sensitivities to the wavelengths of the optical spectrum (see, e.g. CIE, 2015, 2018 and references therein). The canonical definition of luminance was developed for seeing with central vision, that is, for seeing objects whose images are formed in the fovea. The central fovea is composed almost exclusively by L and M cones, with rods and S cones practically absent (ipRGCs do not seem to play a major role in the detection step for this visual task). The standard spectral luminous efficiency function of the human eye in photopic conditions, $V(\lambda)$ (CIE, 1990), is mostly the result of the combined sensitivities of the L and M cones. The luminance, $L_\mathrm{v}$ (units cd m$^{-2}$), can be determined from the spectral radiance by means of the $V(\lambda)$-weighted integral:

$$L_\mathrm{v} = K_m \int_0^\infty V(\lambda)L(\lambda)\mathrm{d}\lambda, \qquad (3)$$

where $K_m$= 683 lm/W is the luminous efficacy of monochromatic radiation of frequency 540 THz (wavelength ~555 nm), $V(\lambda)$ is dimensionless and normalized to 1 at its maximum, and $L(\lambda)$ is expressed in W m$^{-2}$ sr$^{-1}$ nm$^{-1}$. The peculiar value of the $K_m$ factor comes from the fact that luminous and energy units were independently defined by the physicists of the nineteenth century by means of specific experimental setups, and the final connection between both systems of units turned out to require this particular scaling constant.

It is worth noting that the above definition holds for photopic adaptation and central vision. Photopic vision takes place when the eye is fully adapted to luminances higher than about 5 cd m$^{-2}$. For lower luminances the eye sensitivity is better described by the mesopic sensitivity functions $V_{mes,m}(\lambda)$ that take into account the progressively important role of the rod photoreceptors in detecting light, until achieving adaptation levels of 0.005 cd m$^{-2}$, when the vision exclusively depends on the rods, fully within the so-called scotopic regime. A description of the mesopic and scotopic sensitivity functions, as well as of the constants $K_{mes,m}$ that replace $K_m$ for mesopic vision, can be found in CIE (2016). This implies that the appearance of brightness of the light polluted sky will depend, among other factors, on the state of adaptation of the eye. This effect shall be taken into account when using luminance calculations to estimate brightness percepts. Luminances calculated under the assumption of photopic adaptation broadly correspond to the perception of brightness one would have immediately after leaving a normally lit room and going to the dark, before the eyes had time to adapt themselves to the surrounding darkness. Note that the $V_{mes,m}(\lambda)$ functions have overall shapes slightly different from the $V(\lambda)$, and are also spectrally shifted with respect to it. Besides, the central and peripheral retinae have slightly different luminous sensitivities and luminance contrast thresholds, to which any astronomer acquainted with averted vision will certainly attest.

In this paper we analyze the two limiting cases of full photopic and full scotopic adaptation. The luminance under scotopic adaptation is given by an expression analogous to Eq.(3), but with the scotopic luminous efficacy constant $K'_m$= 1700 lm/W instead of $K_m$, and with the spectral luminous efficiency function for scotopic vision $V'(\lambda)$ (CIE, 2010) instead of $V(\lambda)$. The scotopic luminances $L_{v,scot}$ can be easily obtained by multiplying the photopic ones $L_v$ (Eq. 3) by the scotopic-to-photopic luminance ratio (S/P) defined as (Maksimainen et al, 2019):

$$\text{S/P} = \frac{1700 \int_0^\infty V'(\lambda)L(\lambda)\text{d}\lambda}{683 \int_0^\infty V(\lambda)L(\lambda)\text{d}\lambda}, \qquad (4)$$

## 3. MAGNITUDES AND MAGNITUDES PER SQUARE ARCSECOND

Analogously to what happens with the human eye, astronomical instruments have their own spectral sensitivities, as a result of the combined effects of the spectral transmittance of the instrument optics and the detector responsivity to different wavelengths. All these effects can be summarized by the overall spectral sensitivity function of the instrument, $S(\lambda)$, also known as the spectral "$S$ band". By definition, if the spectral irradiance produced by a star on the entrance pupil of an instrument is $E(\lambda)$, the astronomical magnitude of this object in the $S$ band is said to be:

$$mag_S = -2.5 \log_{10} \left( \frac{\int_0^\infty S(\lambda)E(\lambda)d\lambda}{\int_0^\infty S(\lambda)E_r(\lambda)d\lambda} \right), \tag{5}$$

where $E_r(\lambda)$ is an arbitrary (but clearly specified) spectral irradiance that is taken as a refence to set the 'zero point' of the magnitude scale. Traditional choices for $E_r(\lambda)$ are the spectrum of the star Vega (α Lyr) or the absolute (AB) reference of 3631 Jy (Oke, 1974) given by $E_r(\lambda) = 3631 \times (c/\lambda^2) \times 10^{-26}$ W m$^{-2}$ m$^{-1}$, where $c = 2.99792458 \times 10^8$ m s$^{-1}$ is the speed of light in vacuum, $\lambda$ is the wavelength (in m), and the Jy (*jansky*) is a non SI unit of spectral irradiance equivalent to $10^{-26}$ W m$^{-2}$ Hz$^{-1}$. The reader will note here that in the formula for $E_r(\lambda)$ the spectral irradiance is formally expressed as 'per meter' of spectral interval, not per nm, although it is straightforward to convert it to the traditional form by taking into account that 1m=10$^9$ nm. The leading factor of '-2.5' ensures that 100 times less irradiance corresponds to an increase of 5 magnitudes. This is a constant that connects our age with the times of the Hellenistic Greek astronomer Hipparchus (c. 190 - c. 120 BCE) and with the formalization of the magnitude system by Pogson (1856).

Note that the magnitude of any celestial body depends not only on the particular observation band, *S*, but also, and crucially, on the reference irradiance chosen to set the zero-point of the scale. Celestial objects, independently from their linear or angular size (that is, from being 'pointlike' or 'extended'), can be assigned specific magnitudes: they can be determined by measuring their spectral irradiance at the observation site using a sufficiently sensitive spectrometer, or by directly measuring their integrated irradiance with an instrument matched to the *S* band, and applying the equation above.

The brightness of extended objects (e.g. nebulae, galaxies, or the light polluted sky), in turn, is usually described in astrophysics in units of *magnitudes per square arcsecond*, mpsas for short. A small region of the source around a point *P* of an extended object is said to have a brightness of $m_S$ mpsas if each square arcsecond of that region gives rise on the entrance pupil of the telescope (or the eye) to the same irradiance than a $mag_S = m_S$ star would produce. It is easy to see that mpsas are just a way of expressing the in-band radiance arriving to the instrument from *P*. Recall, from equation (2), that the spectral irradiance $E(\lambda)$ produced on a surface by a small angular patch of a source of size $\Delta\omega$ sr, under normal incidence (*z*=0), is $E(\lambda) = L(\lambda)\Delta\omega$, where $L(\lambda)$ is the spectral radiance from *P*. So, the irradiance per unit solid angle, $E(\lambda)/\Delta\omega$, is just $L(\lambda)$. The only caution to apply these formulae is that square arcseconds must be converted to steradians in order to work in consistent SI units. This poses no difficulty since $1\ arcsec^2 = [\pi/(180 \times 60 \times 60)]^2\ \mathrm{sr} = 2.3504 \times 10^{-11}$ sr. Then, we can define the reference radiance $L_r(\lambda)$ for the mpsas scale as the radiance of a source of angular extent $\Delta\omega = 1\ arcsec^2$ that would produce the reference irradiance $E_r(\lambda)$ at the instrument entrance pupil, that is $L_r(\lambda) = E_r(\lambda)/(2.3504 \times 10^{-11}$ sr). From this definition we can directly write for the magnitudes per square arcsecond an equation formally analogous to (5) for the usual magnitudes but now in terms of radiances:

$$m_S = -2.5 \log_{10} \left( \frac{\int_0^\infty S(\lambda)L(\lambda)d\lambda}{\int_0^\infty S(\lambda)L_r(\lambda)d\lambda} \right). \tag{6}$$

## 4. TRANSFORMING MAGNITUDES PER SQUARE ARCSECOND TO LUMINANCES

With the definitions above, the exact transformation between astronomical mpsas and luminance in cd m$^{-2}$ is straightforward. We will keep the generic notation $S(\lambda)$ for the instrument photometric band, although this generic band will be later particularized to the Johnson-Cousins $V$ (Bessell, 1990) not to be confused with the visual $V(\lambda)$ function (CIE, 1990) used in the definition of luminance. First, note that Eq. (6) can be rewritten as

$$\int_0^\infty S(\lambda)L(\lambda)\mathrm{d}\lambda = \left[\int_0^\infty S(\lambda)L_r(\lambda)\mathrm{d}\lambda\right] \times 10^{-0.4 \times m_S}. \tag{7}$$

Multiplying and dividing Eq. (3) by the right- and left-hand terms of equation (7) we get the desired conversion providing the photopic luminance of the night sky, $L_v$, in terms of the brightness in $S$-band mpsas $m_S$ as:

$$L_v = K_m \times \left[\frac{\int_0^\infty V(\lambda)L(\lambda)\mathrm{d}\lambda}{\int_0^\infty S(\lambda)L(\lambda)\mathrm{d}\lambda}\right] \times \left[\int_0^\infty S(\lambda)L_r(\lambda)\mathrm{d}\lambda\right] \times 10^{-0.4 \times m_S}, \tag{8}$$

where $L_v$ has units of cd m$^{-2}$, $K_m$= 683 lm W$^{-1}$, and $L_r(\lambda)$ is expressed in W m$^{-2}$ sr$^{-1}$ nm$^{-1}$. Equation (8) can be conveniently shortened as:

$$L_v = L_0 \times 10^{-0.4 \times m_S}, \tag{9}$$

where $L_0$, of units cd m$^{-2}$, also known as the photopic zero-point luminance, is the product of the first three factors of the right-hand side of equation (8).

Note that for any given photometric band $S(\lambda)$ with a well specified reference radiance $L_r(\lambda)$, the first and third factors of equation (8) are constant, whereas the second one (the ratio of the $V(\lambda)$ to $S$ in-band radiances) depends on the spectral composition of the incident light, $L(\lambda)$. Only in the very particular case when the photometric band $S(\lambda)$ strictly coincides with the CIE visual one, $V(\lambda)$, the second factor equals 1, and $L_0$ turns out to be independent from the spectrum of the light (Bará, 2017). Then, the zero-point luminance $L_0$ that determines the $S$-band mpsas to cd m$^{-2}$ conversion is not a single value but must be calculated for each $L(\lambda)$ on a case by case basis. However some general insights on the order of magnitude of $L_0$ can be devised by analysing actual skyglow spectra. This is the subject of the next section.

## 5. ZERO-POINT LUMINANCES IN THE $V$ BAND FOR SPECTRA RECORED AT DIFFERENT SITES

The zero-point luminances $L_0$ for a set of skyglow spectra obtained at different sites under different atmospheric and moonlight conditions are reported in this section. The photometric $S$ band will be assumed to be the Johnson-Cousins $V$ according to the Bessell (1990) definition, and the luminance $L_v$ will be calculated firstly under the assumption of photopic adaptation. Note that for the Johnson-Cousins *V* the reference AB in-band radiance is equal to:

$$\left[\int_0^\infty S(\lambda) L_{r,AB}(\lambda) d\lambda \right] = 137.8716 \quad \text{Wm}^{-2}\text{sr}^{-1} \tag{10}$$

whereas in case of using the Vega's spectral power distribution, $L_{r,VEGA}$, to set the scale of the *V* magnitudes some additional choices have to be made. For this work we have defined the Vega scale such that the Johnson-Cousins *V* magnitude of Vega is +0.03, using the Vega irradiance spectrum quoted by Rieke et al. (2008), available at VizieR (http://vizier.u-strasbg.fr/viz-bin/VizieR?-source=J%2FAJ%2F135%2F2245). The reference radiance for this scale turns out to be 146.7728 W m$^{-2}$ sr$^{-1}$, that is, 1.0646 times higher than the AB one in equation (10). This proportionality factor will relate the values of the zero point luminances $L_0$ (AB) and $L_0'$ (Vega) to be used in equation (9), depending on the chosen scale.

The first 19 cells of Figure 2 show the sky spectra analysed in this section. Seventeen out of these nineteen spectra were obtained using the Spectrometer for Aerosol Night Detection (SAND) (Aubé, 2008; Hänel et al, 2018) at Los Cancajos (La Palma, Canary Islands), El Leoncito Observatory (Argentina), Hefei (China), Hong Kong (China), Linan (China), Montsec Observatory (Catalunya), Teide Observatory (Tenerife, Canary Islands), Roque de los Muchachos Observatory (La Palma, Canary Islands), Universitat de Barcelona (Catalunya), and Universidad Complutense de Madrid (Spain). The spectral resolution of the SAND-4 is 2.4 nm determined from the FWHM=2.8 nm. The remaining two spectra correspond to Tucson (AZ, USA) and Calar Alto Observatory (Almería, Spain) (Sánchez et al, 2007). The Tucson spectrum was acquired with a custom-built spectrograph that gives a spectral resolving power (R) of about 500, and the image was captured with a ZWO AS1294 CCD camera. The device has a 50 μm x 3 mm slit. It is fed by a 50 mm focal length lens, for projected dimensions on the sky of 200 arcseconds by 3.4 degrees. All spectra were obtained toward the zenith, excepting El Leoncito clear A, that was taken at 15 deg elevation and 160 deg azimuth, and Tucson, obtained 22 km northwest of the city center, with the lens feeding the spectrograph pointed 20 degrees above the horizon toward the azimuth corresponding to the brightest part of the skyglow distribution over the city. The spectra were taken on moonless nights, excepting one from El Leoncito Observatory ("clear, Moon"). The acquisition dates and times are indicated in the second column of Table 1. Recall that the natural brightness of the sky, including extra-atmospheric sources and airglow emission lines, does change over different time scales, from long term (e.g. solar cycle) to short-term ones within the same night. The same can be said about the artificial component of the skyglow and its basic constituents, whose emissions vary noticeably over the course of the night depending on the amount of light used at each time interval (Dobler et al, 2016; Bará et al. 2019). The spectra provided here shall hence be taken as particular realizations of the spectral radiance at each site, and not as features unequivocally characterizing the site sky.

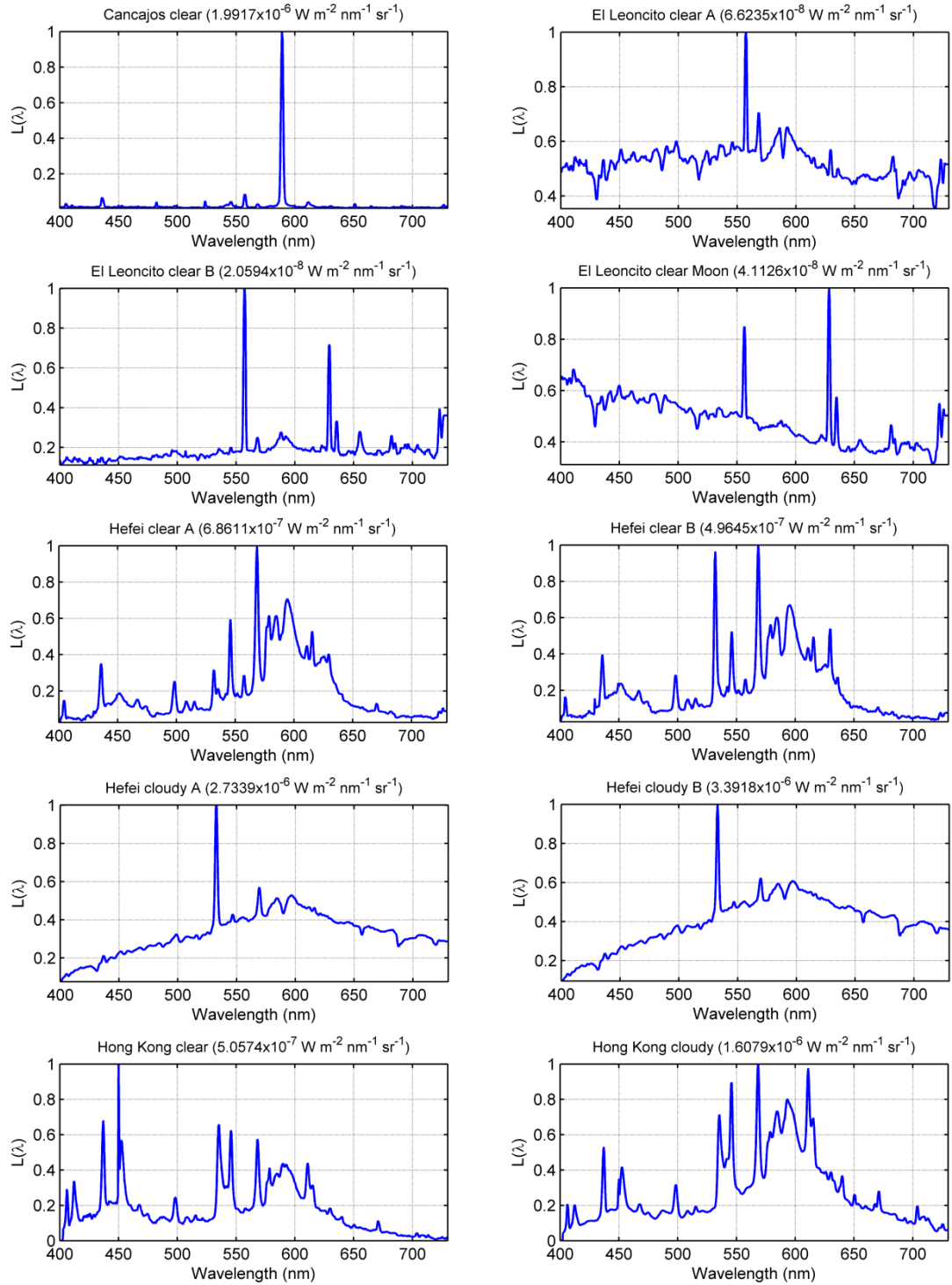

**Figure 2.** Samples of 19 skyglow radiance spectra at selected locations throughout the world. See text and Table 1 for details. The spectra are normalized to 1 at their maxima for display purposes. The absolute peak value of each spectrum, in W m$^{-2}$ nm$^{-1}$ sr$^{-1}$ is indicated in the caption of each curve. The lower right cell ("Lamps") displays the spectra of typical high pressure sodium (blue line, nominal CCT 2000 K), and light-emitting diode lamps (black and gray lines, nominal LED CCTs 2400, 3000, 4000, and 5000 K). In LED, higher CCTs are associated with higher values of the blue-to-yellow peak ratio.

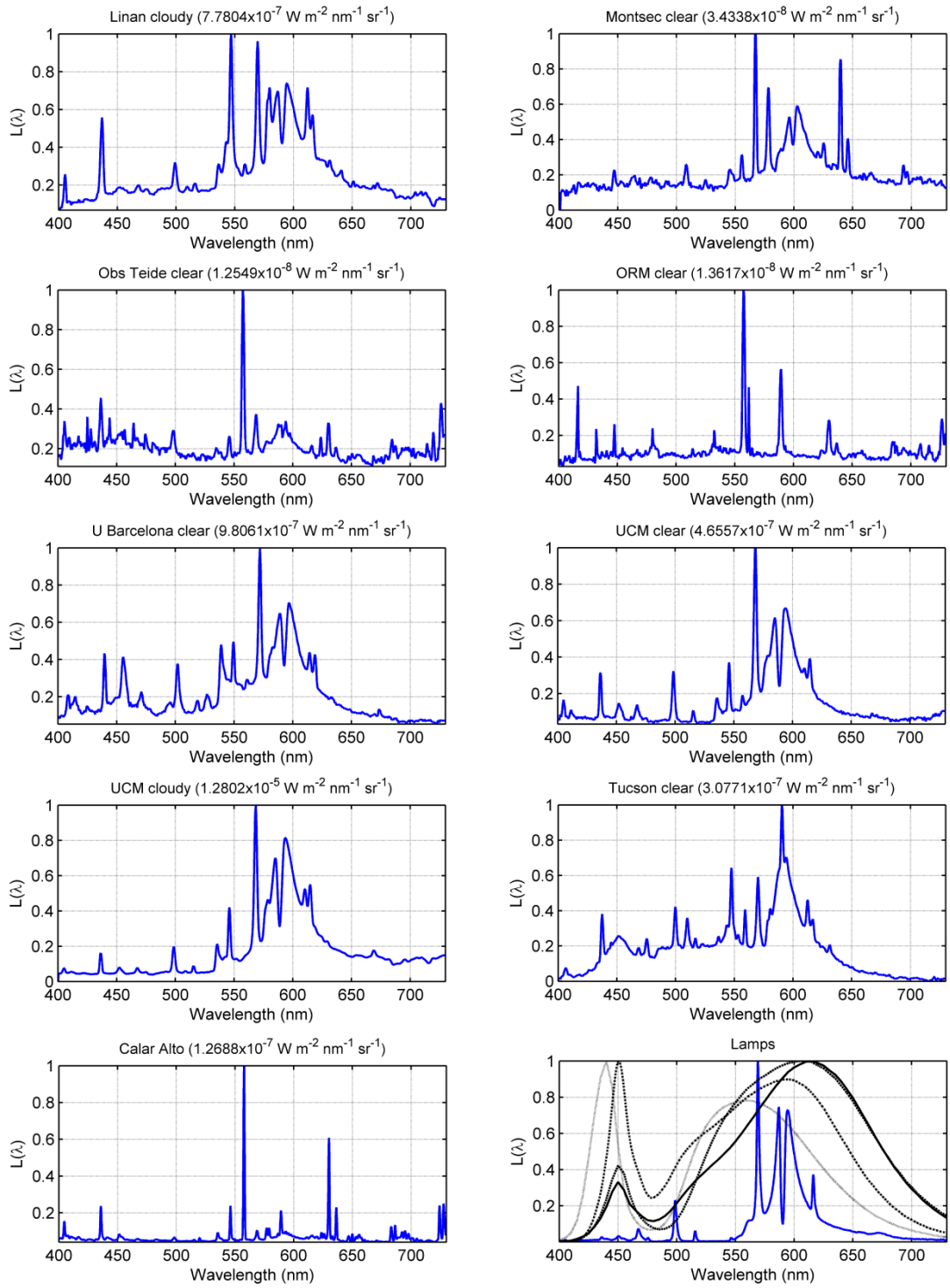

**Figure 2.** (cont.)

The spectra shown in Figure 2 are normalized to unity at their maxima only for display purposes. The absolute peak value of each radiometrically calibrated sky spectrum, in W m$^{-2}$ nm$^{-1}$ sr$^{-1}$, is indicated in the caption of each curve. For computing the value of the second factor of the right-hand side of equation (8), which is a ratio of weighted integrals of $L(\lambda)$, the precise units in which the spectra are expressed are of no concern as far as they are linear, since they cancel out when taking the ratio. This fact would allow one to use arbitrary linear units for $L(\lambda)$, and even spectral data acquired with spectrometers without a precisely calibrated field-of-view, what strongly simplifies the calculations. The last cell of Figure 2 displays the spectra of a typical high pressure sodium (HPS, nominal CCT 2000 K), and several light-emitting diode (LED, nominal CCTs 2400, 3000, 4000, and 5000 K) lamps from two different spectral libraries, the LSPDD database (Sánched de Miguel et al, 2017) and the LICA UCM spectra database (Tapia et al, 2019), whose zero-point luminances have also been calculated and are displayed in Figures 3 and 5 for comparison.

Table 1 summarizes the values of the zero-point luminances in equation (9) for the sky spectra, in the AB ($L_0$) and Vega ($L_0'$) scales of the Johnson-Cousins $V$ photometric band. The actual sky brightnesses at the observation sites, calculated from the calibrated radiance spectra using equation (6) for the Johnson-Cousins $V$-band, in AB mpsas, are listed in the third column. The sky brightnesses in μcd m$^{-2}$ for potential skies with these spectral compositions and m$_V$=22.0 mpsas in both scales are listed in columns 6 and 7. The approximate values of the Correlated Color Temperatures (CCT), calculated according to McCamy (1992) are listed in column 8. The last column of this Table displays the S/P ratio of each spectrum, calculated according to equation (4).

Table 1. Johnson-Cousins m$_V$ to cd m$^{-2}$ zero-point luminances for the sky spectra shown in Figure 2

| Site | Date and time (UTC) | Sky Brightness m$_V$ (AB) | $L_0$ (AB) | $L_0'$ (Vega) | Photopic luminance for m$_V$=22.0 mpsas | | CCT | S/P ratio |
|---|---|---|---|---|---|---|---|---|
| | | | | | AB | Vega | | |
| | yyyymmdd:hhmm | mpsas | in units 10$^5$ cd m$^{-2}$ | | μcd m$^{-2}$ | | K | |
| Cancajos clear | 20100311:0037 | 18.54 | 1.3166 | 1.4016 | 209 | 222 | 2727 | 1.1005 |
| El Leoncito clear A | 20140111:0217 | 19.04 | 1.1302 | 1.2032 | 179 | 191 | 5314 | 2.1814 |
| El Leoncito clear B | 20140111:0533 | 21.44 | 1.1609 | 1.2359 | 184 | 196 | 4216 | 1.8454 |
| El Leoncito clear Moon | 20140114:0506 | 19.70 | 1.1107 | 1.1824 | 176 | 187 | 6728 | 2.4721 |
| Hefei clear A | 20131226:1243 | 17.32 | 1.2799 | 1.3626 | 203 | 216 | 2730 | 1.1309 |
| Hefei clear B | 20140122:1105 | 17.60 | 1.2373 | 1.3172 | 196 | 209 | 3133 | 1.3081 |
| Hefei cloudy A | 20140118:1557 | 15.36 | 1.1504 | 1.2247 | 182 | 194 | 3920 | 1.7262 |
| Hefei cloudy B | 20140118:1656 | 14.94 | 1.1508 | 1.2251 | 182 | 194 | 3902 | 1.7329 |
| Hong Kong clear | 20160624:1255 | 17.77 | 1.1869 | 1.2636 | 188 | 200 | 4641 | 1.7579 |
| Hong Kong cloudy | 20160613:1307 | 16.00 | 1.2345 | 1.3142 | 196 | 208 | 3128 | 1.3062 |
| Linan cloudy | 20111114:1104 | 16.90 | 1.2282 | 1.3074 | 195 | 207 | 3244 | 1.3704 |
| Montsec clear | 20160503:2101 | 20.73 | 1.2611 | 1.3425 | 200 | 213 | 2830 | 1.4406 |
| Obs Teide clear | 20100210:0057 | 21.92 | 1.1734 | 1.2492 | 186 | 198 | 5548 | 2.1470 |
| ORM clear | 20100311:2240 | 22.34 | 1.1475 | 1.2216 | 182 | 194 | 4918 | 1.8834 |
| U Barcelona clear | 20160930:2134 | 16.84 | 1.2299 | 1.3093 | 195 | 208 | 3417 | 1.4354 |
| UCM clear | 20131105:2351 | 18.08 | 1.3211 | 1.4064 | 209 | 223 | 2586 | 1.0020 |
| UCM cloudy | 20140402:1959 | 14.33 | 1.3436 | 1.4303 | 213 | 227 | 2248 | 0.8198 |
| Tucson clear | 20191023:0409 | 18.09 | 1.1838 | 1.2602 | 188 | 200 | 3750 | 1.6378 |
| Calar Alto | 2005-2006(*) | 20.57 | 1.1840 | 1.2605 | 188 | 200 | 4256 | 1.7590 |

(*) Sanchez et al (2007)

The values of the photopic zero-point luminances of these spectra, in the AB and Vega scales, are represented versus the CCT in Fig. 3a. The zero-point luminances corresponding to blackbody radiation with CCT comprised between 1500 K and 7500 K (Bará, 2019) are also shown for comparison. In Figure 3b we show for completeness the corresponding scotopic zero-point luminances, that is, the zero points $L_0$(scotopic) to be used in equation (9) to transform the Johnson-Cousins *V*-band mpsas to the luminances that would be perceived by a human eye under fully dark adaptation (see Section 2). These values are obtained by following the same procedure outlined in Section 4, but using the scotopic $V'(\lambda)$ and $K'_m$ instead of the photopic $V(\lambda)$ and $K_m$, respectively. This is equivalent to multiplying the photopic values by the corresponding S/P ratio. The reader may notice that $L_0$(scotopic) increases with CCT, a behaviour opposite to that shown by the photopic $L_0$. This reflects the basic fact that, for most observable spectra, an increase in the CCT, keeping constant the integrated radiance within the Johnson-Cousins *V*-band and hence the *V*-band magnitude, tends to provide an increased integrated radiance within the scotopic $V'(\lambda)$, whose passband is shifted towards the blue relative to the Johnson-Cousins *V*, whereas the opposite effect is observed in the photopic $V(\lambda)$ band, slightly shifted to the red. In other words, for any constant value of the Johnson-Cousins *V*-band magnitude, higher CCT (bluer) spectra tend to increase the visual efficacy of the radiation for eyes scotopically adapted, while reducing it for photopically adapted ones. The larger dynamic range of the $L_0$(scotopic) values in comparison with the photopic ones is due to the spread in the corresponding scotopic-to-photopic ratios (S/P) defined in equation (4) and listed in Table 1.

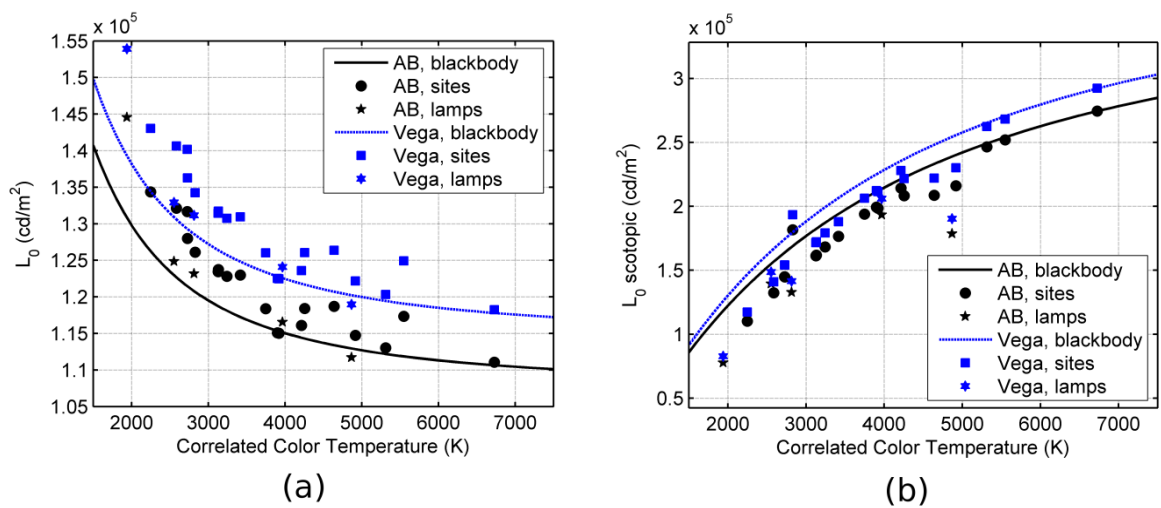

**Figure 3.** Values of the zero-point luminances (cd m$^{-2}$) for (a) photopic adaptation, (b) scotopic adaptation. In both figures: (solid line) blackbody spectra of temperature 1500 K to 7500 K using the AB scale for m$_V$; (dotted line) blackbody spectra, using the Vega scale; (circles) skyglow spectra of Figure 2, using the AB scale; (squares) skyglow spectra of Figure 2, using the Vega scale; (pentagrams) typical spectra of the direct radiance from HPS (nominal CCT 2000 K) and LED (nominal CCT 2400, 3000, 4000, and 5000 K) lamps in the AB scale; (hexagrams) idem, in the Vega scale. The actual CCT of each lamp may vary slightly from the nominal one indicated by the manufacturer, due to allowed industrial tolerances.

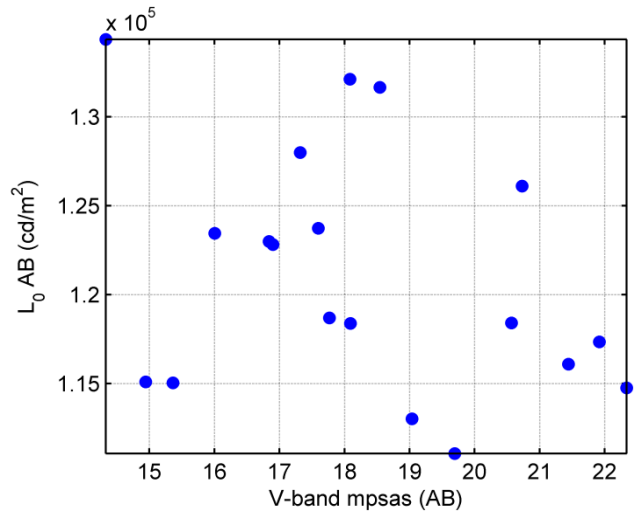

**Figure 4.** Zero-point luminance vs actual brightness of the observing sites in the Johnson-Cousins *V*-band (AB)

Figure 4 shows the values of the photopic AB zero-point luminances as a function of the actual brightness of the observing sites in the *V* band (AB mpsas), deduced from the radiometrically calibrated spectra by application of equation (6). The points are considerably scattered, with no clear correlation, although values for dark skies ($m_V$>21.0 mpsas) tend to lie at the lower end of the range. Figure 5 displays the S/P ratio of the measured spectra, as a function of their CCT. The sky spectra S/P ratio are in the range 0.8-2.5 and tend to increase with CCT, similar to the behaviour observed for common spectra of lamps (Fotios & Yao, 2018).

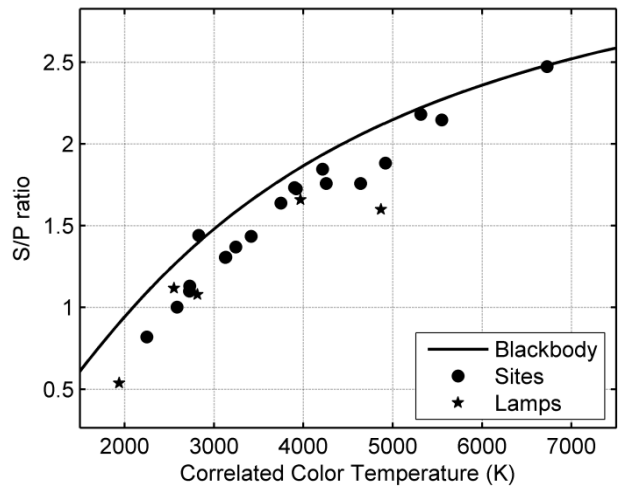

**Figure 5.** S/P ratio vs CCT of the measured sky spectra (circles) and direct radiance of typical HPS (nominal CCT 2000 K) and LED (nominal CCT 2400, 3000, 4000, and 5000 K) sources (pentagrams). The values for blackbody spectra with CCT between 1500 K and 7500 K are also shown for comparison (solid line).

## 6. DISCUSSION

The examples above show that the zero-point luminances required to transform the Johnson-Cousins *V* mpsas into luminance in cd m$^{-2}$ are dependent on the shape of the skyglow spectrum. The photopic skyglow luminance corresponding to m$_V$=0.00 is found to vary between 1.11-1.34 x 10$^5$ cd m$^{-2}$ if the m$_V$ are reported in the absolute (AB) magnitude scale ($L_0$), and between 1.18-1.43 x 10$^5$ cd m$^{-2}$ if the Vega scale with mag$_V$(Vega)=+0.03 is used instead ($L_0'$). These values are higher than the one commonly encountered in the literature to perform this transformation, which is 1.08 x 10$^5$ cd m$^{-2}$. As discussed elsewhere (Bará, 2017) the latter value would apply only for transforming *V* band mpsas to cd m$^{-2}$ if the spectrum corresponds to a blackbody source with effective temperature 9550 K, and the mpsas are measured in the Johnson-Cousins *V* band with a radiance zero point of 136.8 W m$^{-2}$ sr$^{-1}$, slightly lower than the AB value used in the present work.

From Fig. 3a it can be seen that the values of the photopic constants for actual skyglow spectra tend to be higher than the ones that would correspond to blackbody radiation of comparable CCT. In both cases higher CCT values are associated with smaller values of the zero-point luminances. Fig. 3a also provides a partial answer to the question of whether the CCT can be a good proxy for guessing the value of the zero-point luminance of skyglow. Whereas blackbody radiators show a precise and well-behaved dependence of $L_0$ on absolute temperature, skyglow spectra give rise to a considerably more scattered pattern when represented against CCT. However, if only a modest degree of precision is required, skyglow photopic zero-point luminances can be roughly estimated to be 0 to 0.1 x 10$^5$ cd m$^{-2}$ higher than the ones for blackbody radiation of the same CCT. The existence of some degree of correlation between the zero-point luminances and the CCT associated with the different spectra could be anticipated, since the CIE $V(\lambda)$ function is embedded in the calculation of the CCT, although in a non-linear way. On the other hand, the scotopic zero-points tend to behave in the opposite way (Fig. 3b).

An issue that has been subject of some debate is which luminance level should be assigned to skies of m$_V$=22.0 mpsas, deemed to correspond to the average zenithal brightness of the natural night sky (Pilachowski et al, 1989; Leinert, 1998; Patat, 2008). As shown in Table 1, the photopic visual luminance of potential skies of m$_V$=22.0 with our measured spectra is comprised between 176 and 213 µcd m$^{-2}$ (m$_V$ in AB), or between 187 and 227 µcd m$^{-2}$ (m$_V$ in Vega), depending on sites. For the darkest observatories these values tend to concentrate in the lower end of the interval. Note however that the overall correlation between the zero point luminance and the sky brightness in Johnson-Cousins *V* mpsas at the different observation sites (Fig 4) is far from being strong: skies with very different brightness may present similar zero-point luminances.

The lack of a perfect correspondence between Johnson-Cousins m$_V$ and cd m$^{-2}$, as is expected when comparing brightness metrics obtained in different photometric bands, strongly suggests the convenience of always reporting night sky brightness data in the native observational band in which they were gathered. Conversions to other bands can of course be done for the readers' benefit, clearly stating the assumptions made to establish the transformations. Note that this is a relatively easy process to carry out in the forward direction (that is, from the native measurements to their approximate equivalence in other band), but it is sometimes quite difficult to perform in the backward direction (from the transformed values to the native ones) in absence of extensive explanations that are not always found (or at least, not with sufficient level of detail) in the corresponding literature.

The numerical results presented in Section 5 of this work focus on the transformations from the Johnson-Cousins *V* band to visually relevant outcomes, due to the pervasive use of that band in skyglow studies and light pollution research. The general conversion approach described in Section 4, however, can be applied to any photometric band of astrophysical interest. A relevant comment from the reviewer elicits the issue of whether other band choices -namely, the widely used g' filter (Fukugita et al., 1996)- could be beneficial for light pollution research, in order to get a stronger signal from the atmospheric scattering of the blue peak of phosphor-coated LED sources while avoiding at the same time some natural airglow emission lines. This is an interesting proposal that deserves careful consideration and should be addressed in future works.

# 7. CONCLUSIONS

The visual brightness of the night sky, in luminance units of cd $m^{-2}$, is not a single-valued function of the brightness in magnitudes per square arcsecond (mpsas) in other photometric bands, because the conversion depends on the spectral power distribution of the skyglow. We analysed in this paper the trasnformation between mpsas and luminance, in order to provide a rigorous assessment of the luminance associated with the visual perception of the night sky. To that end, we calculated the photopic and scotopic zero-point luminances for a set of skyglow spectra recorded at different places in the world, including strongly light-polluted locations as well as sites with nearly pristine, naturally dark night skies. The photopic skyglow luminance corresponding to $m_V$=0.00 in the Johnson-Cousins *V* band is found to vary between 1.11-1.34 x $10^5$ cd $m^{-2}$ if $m_V$ is reported in the absolute (AB) magnitude scale, and between 1.18-1.43 x $10^5$ cd $m^{-2}$ if a Vega scale for $m_V$, with $mag_V$(Vega)=+0.03, is used instead. The resulting photopic visual luminance of the sky for $m_V$=22.0 mpsas ranges between 176 and 213 µcd $m^{-2}$ (AB), or 187 and 227 µcd $m^{-2}$ (Vega). The corresponding scotopic values can be obtained by multiplying the photopic ones by the individual S/P ratio of the spectra, that take values in the range 0.8 to 2.5. The dependence of the transformation constants on the CCT of the spectral power distributions is also reported.

## ACKNOWLEDGEMENTS

This work was supported by Xunta de Galicia/FEDER, grant ED431B 2017/64 (SB). JZ acknowledges the support from ACTION, a project funded by the European Union H2020-SwafS-2018-1-824603, and RTI2018-096188-B-I00. MA research activities were supported by the Fond de recherche du Québec, Nature et Technologies (FRQNT). The authors thank Scott Tucker for providing the Tucson night sky spectrum. Thanks are also due to the reviewer for useful suggestions and comments.